\documentclass{emulateapj}

\def\msun{{\rm ~M}_{\odot}}

\def\mpy{{\rm ~M}_{\odot} {\rm ~yr}^{-1}}

\begin{document}

\title{Rates and Delay Times of Type Ia Supernovae}

 \author{Ashley J. Ruiter\altaffilmark{1,2}, Krzysztof
 Belczynski\altaffilmark{3,4}, Chris Fryer\altaffilmark{3}}
         
 \affil{
     $^{1}$ New Mexico State University, Dept of Astronomy,
        1320 Frenger Mall, Las Cruces, NM 88003\\
     $^{2}$ Harvard-Smithsonian Center for Astrophysics, 
	60 Garden St., Cambridge, MA 02138 (Predoctoral Fellow)\\
     $^{3}$ Los Alamos National Laboratory, Los Alamos, NM 87545\\
     $^{4}$ Oppenheimer Fellow\\
     aruiter@nmsu.edu, kbelczyn@nmsu.edu, clfreyer@lanl.gov}

\begin{abstract} 

We analyze the evolution of binary stars to calculate synthetic rates
and delay times of the most promising Type Ia Supernovae progenitors.
We present and discuss evolutionary scenarios in which a white dwarf
reaches the Chandrasekhar-mass and potentially explodes in a Type Ia
supernova.  We consider: Double Degenerate (DDS; merger of two white
dwarfs), Single Degenerate (SDS; white dwarf accreting from H-rich
companion) and AM Canum Venaticorum (AM CVn; white dwarf accreting from
He-rich companion) scenarios. The results are presented for
two different star formation histories; burst (elliptical-like
galaxies) and continuous (spiral-like galaxies). 
It is found that delay times for the DDS in our standard model
(with common envelope efficiency $\alpha_{\rm CE} = 1$) follow a power-law 
distribution.  For the SDS we 
note a wide range of delay times, while AM CVn progenitors produce 
a short burst of SNe Ia at early times.  
The DDS median delay time falls between $\sim 0.5 - 1 $ Gyr; the SDS 
between $\sim 2 - 3$ Gyr; and the AM CVn between $\sim 0.8 - 0.6$ Gyr
depending on the assumed $\alpha_{\rm CE}$.  
For a Milky Way-like galaxy we estimate the rates of SNe Ia arising from
different progenitors as: $\sim 10^{-4}$ yr$^{-1}$ for the SDS and AM CVn, and
$\sim 10^{-3}$ yr$^{-1}$ for the DDS. We point out that only the rates for two
merging carbon-oxygen white dwarfs, the only systems found in the DDS, are
consistent with the observed rates for typical Milky Way-like spirals.
We also note that DDS progenitors are the dominant population in 
elliptical galaxies.  The fact that the delay time distribution for the DDS 
follows a power-law implies more Type Ia supernovae 
(per unit mass) in young rather than in aged populations.
Our results do not exclude other scenarios, but strongly indicate that the DDS 
is the dominant channel generating SNe Ia in spiral galaxies, at least in the 
framework of our adopted evolutionary models.  
Since it is believed that white dwarf mergers cannot produce a thermonuclear explosion
given the current understanding of accreting white dwarfs, either the 
evolutionary calculations along with accretion physics are incorrect, or the 
explosion calculations are inaccurate and need to be revisited.  

\end{abstract}


\section{Introduction}

Type Ia Supernovae (SNe Ia) play an important role in astrophysics as
cosmological distance indicators.  Additionally, they provide iron
peak elements, having direct consequences for the chemical evolution
of galaxies \citep{RPK95,MG86,DV03}.  Currently $\gtrsim 2000$ SNe Ia
have been observationally
confirmed\footnote{http://www.cfa.harvard.edu/iau/lists/Supernovae.html},
some as distant as $z \sim 1.55$ \citep{Str04}.  Empirically-derived
relationships between light curve properties and intrinsic luminosity
(i.e., $\Delta m_{15}$ \citep{Phi93} and stretch-factor $s$,
\citep{Per97}) have made it possible to `standardize' absolute
magnitudes of SNe Ia light curves over a wide variety of host galaxy
environments.\footnote{It has been shown that SNe Ia originating among
  young stellar populations are overall more luminous than those
  associated with older stellar populations \citep[i.e.,][]{Ham95}.}
Their use as `standard candles' on cosmological scales has led to the
realization that the expansion rate of the universe is accelerating,
and has enabled accurate estimations of $\Omega_{\Lambda}$ and
$\Omega_{M}$ \citep[e.g.,][]{Sch98,Rie98,Per99}.  However, using SNe
Ia in order to set the distance scale for the determination of
cosmological quantities requires the (still unfounded) assumption that
the physical properties of their progenitors are unchanging with
redshift.

Despite the continued use of SNe Ia as standard candle distance
indicators, their origin remains uncertain.  It is generally accepted
that SNe Ia arise from the total disruption of a Chandrasekhar mass
($\sim 1.4$ M$_{\odot}$) carbon-oxygen white dwarf (WD) as a 
result of thermonuclear explosion.
This hypothesis is supported by the fact that the amount of energy
observed in the explosions ($\sim 10^{51}$ erg; \citet{Thi04}) is
equal to the amount which would be produced in the conversion of
carbon and oxygen into iron \citep[see e.g.,][for a review]{Liv00}.
It is natural to presume that the exploding white dwarf must accrete
matter from a close stellar companion until reaching the critical
Chandrasekhar mass, though the nature of the companion, the rate and
efficiency at which mass is accumulated onto the white dwarf, and
which array of conditions are necessary in order for the white dwarf
to ignite explosively, are not well understood \citep{NIK97}.

In order to constrain the nature of SN Ia progenitors, the SN Ia rate
has been studied as a function of parent galaxy stellar mass, star
formation rate, colour, morphology, and radio power by several groups
\citep[e.g.,][]{Man05,Sul06,CM06,DP03}.  \citet{Man05} found that the
SN Ia rate is higher in bluer, later type galaxies, supporting the
hypothesis that there is a non-negligible number of SNe Ia originating
from young progenitors.  \citet{Sul06} found that the SN Ia rate per
unit mass increases as a function of star formation activity, a trend
which coincides with the results of \citet{CM06}, who used chemical
evolution models to derive the rate of SNe Ia as a function of galaxy
Hubble type.  \citet{DP03} discovered an enhanced SN Ia rate in
radio-loud galaxies compared to their radio quiet counterparts,
possibly as a result of past interactions/mergers with dwarf galaxy
companions leading to an increased number of newly-formed or captured
young stars (see also \citet{Del05}).  The evolution of the supernova 
rate (both core-collapse
and Type Ia) as a function of cosmic time was investigated by
\citet{MDP98}, who convolved a set of theoretical characteristic SN
delay times with the cosmic star formation history, resulting in an
estimate of SN rates out to intermediate redshifts.  Later studies
showed that a single-component delay time could not be reconciled with
the observed mass of iron in galaxy clusters and the corresponding
ratio of core-collapse to Ia SN rates \citep{MG04}.  

SNe Ia are observed in 
both young and old galaxies \citep{BV93}, thus it is 
natural to presume that the progenitors may originate from 
both young and old stellar populations.  Recently, it has 
been found that SNe Ia appear to span a wide range 
of delay times which is bimodal in nature, consisting of 
a `prompt' population with short delay times, and a `tardy' 
population whose average delay time distribution (DTD) is much 
wider and is best described by a decaying 
exponential function \citep{SB05,MDP06,Dil08}.  
However, whether the apparent bimodal DTD shape is limited 
to low-redshift SNe alone or whether it also applies to 
SNe at higher redshift is still unclear \citep{DSR08}. 

Two formation scenarios have emerged as the most likely  
channels for SNe Ia progenitors: The Single Degenerate Scenario 
\citep[SDS,][]{WI73,Nom82} and the Double 
Degenerate Scenario \citep[DDS,][]{IT84,Web84}.  
The SDS is encountered when a WD accretes H-rich matter during stable 
Roche Lobe Overflow (RLOF) from a stellar companion; either a main sequence
or an evolved (giant) star. The WD  
increases in mass up to the Chandrasekhar mass limit, enabling 
carbon to ignite explosively in the WD center causing a SN Ia.
The DDS is the result of a merger 
of two white dwarfs.  If the combined mass of the merger exceeds the 
Chandrasekhar mass, the result may be a SN Ia.
Additionally it has been suggested that a third channel, the AM CVn 
channel\footnote{SWB-like in \citet{BBR05}.}, may account for 
1 \% of SNe Ia \citep{SY05}.  
AM CVn binaries \citep[see e.g.,][]{Nel05,War95} are ultra-compact 
systems involving a WD accretor and a helium-rich donor exchanging 
matter via RLOF.  The donors are expected to be small stars 
given the small orbital size (close orbits; $\lesssim 1$ hr).
Other possible SNe Ia formation scenarios have been proposed, 
though they likely do not account for the majority of SNe Ia 
(e.g., sub-Chandrasekhar mass SNe Ia \citep{WW94}, 
common envelope WD mergers \citep{LR03,App91,SS74}). 

In this study we follow the evolution of stellar populations in two
different environments, that which is typical for an
elliptical galaxy (instantaneous burst of star formation at $t=0$) and that 
typical for a spiral galaxy (continuous star formation).  
We also employ two different parameterizations for common envelope 
(CE) evolution.  
To substantiate our conclusions, we use exactly the same total stellar 
mass and metallicity for each population; the only differences in our
model galaxies are the star formation histories, and the assumed CE 
removal efficiency $\alpha_{\rm CE}$.  We show which SN Ia progenitors are the most 
likely (from an evolutionary perspective) for each host galaxy type, and 
derive delay times and rates for the most promising SNe Ia progenitor scenarios.  
We compare our results to those of previous studies and we discuss our results 
(e.g., in terms of explosion physics) in the last two sections, respectively.  

\section{Model Description}

Our stellar evolution calculations are performed using the {\tt StarTrack} population 
synthesis code.  A detailed description of the input physics is presented in 
\citet{BKB02,Bel08}.  
Single star evolution is followed from the ZAMS until 
remnant formation employing modified analytic formulae and
evolutionary tracks from \citet{HPT00}.  Evolution of binary stars is 
more complex, and several processes important for field binary evolution are accounted 
for, such as tidal interactions, mass transfer phases, common envelope
evolution, supernova kicks, magnetic braking and gravitational
radiation \citep[see][for formulae]{Bel08}.  
We incorporate recent prescriptions for mass growth of WDs, employing 
accretor mass-dependent accumulation efficiencies which may lead to 
nova explosions, stable burning, or optically thick WD winds 
\citep[][see below]{Nom07,KH04,HKN99,KH99,PK95,Has86}.  The physical properties of
the stars are computed throughout the evolution.  

A merger between two white dwarfs may lead to a DDS SN Ia.  
A detached WD-WD binary will eventually reach contact due to angular 
momentum loss from the emission of gravitational radiation, 
and if the binary configuration (e.g., mass ratio) leads to a merger, the less massive 
WD is accreted onto the more massive WD \citep[see e.g.,][]{TY79}.  
If the combined mass of the merger exceeds the Chandrasekhar mass, it 
is recorded as a potential DDS SN Ia progenitor.  We assume a 
priori that {\em every} WD-WD merger with $M \geq 1.4$ M$_{\odot}$
consisting of CO-CO, CO-He, or He-He WDs leads to an 
instantaneous SN Ia.\footnote{In \citet{BBR05}, it was assumed that
  a merger between any two WDs, including ONe WDs, with a combined mass exceeding 1.4
  M$_{\odot}$ would lead to a SN Ia, though CO-CO mergers made up $88$
  \% of the DDS SNe Ia in that study.}  
Mergers involving other WD types occur in our 
simulations but we do not count them as SN Ia progenitors.    

A SDS or AM CVn SN Ia may result from the accumulation of matter on a
white dwarf's surface via stable RLOF from a stellar companion.  
For accretion of hydrogen-rich material,  
strong nova explosions inhibit the accumulation
of hydrogen on the WD surface for very low mass transfer rates $<10^{-11}$
yr$^{-1}$ \citep{PK95}.  For mass transfer rates above this threshold, 
we interpolate over the results of \citet{PK95} to obtain the
mass accretion efficiencies, and we account for optically thick winds at high mass 
transfer rates \citep[][see also Belczynski, Bulik \& Ruiter (2005)
  Section 2.3 for a more detailed description]{HKN99}.  
The only difference between the accumulation efficiencies in this work
and those of \citet{BBR05} is that here we have
additionally included an updated prescription for accretion of hydrogen-rich 
matter from \citet{Nom07}, in which fully efficient accumulation is 
only achieved for a very narrow range of mass transfer rates, 
and is also dependent upon the white dwarf accretor mass
\citep[][see equations 5 \& 6]{Nom07}.  

Accretion of helium-rich matter is treated in the same fashion as in
\citet{BBR05}: accretion prescriptions are adopted from \citet{KH99}.  
However this study contains one major difference which affects
our results: though in this work we do allow for the formation of
sub-Chandrasekhar mass SNe Ia, we do not include these binaries
as potential SN Ia progenitors here.  In \citet{BBR05}, a large fraction ($61$ \%) of the
Type Ia SNe which contributed to the presented delay times in the standard model were in
fact {\em sub-Chandrasekhar} SNe Ia, in which the accumulation of $\sim 0.1$ M$_{\odot}$ of
He-rich material on the WD surface could lead to an edge-lit
detonation and subsequent SN Ia \citep{WW94,KH99}.  

The criteria used here for defining SNe Ia progenitors arising from different 
formation channels are different from the work of \citet{BBR05}.  
In this work, we only consider accreting WDs which have obtained a mass of 
1.4 M$_{\odot}$ as potential SNe Ia. 
We make the distinction between SN Ia progenitors with CO white dwarfs 
accreting from non-degenerate hydrogen-rich companions (SDS), 
and white dwarfs accreting from helium-rich companions (AM CVn).   
We note that He WDs never reach the Chandrasekhar mass in our simulations 
with the adopted accumulation physics, and we have assumed that oxygen-neon-magnesium 
WDs collapse to form a neutron star (accretion induced collapse) upon reaching 
the Chandrasekhar mass, rather than producing a SN Ia \citep[e.g.,][]{Miy80}.  
For the SDS and AM CVn cases, we record the binary as a SN Ia 
once the accreting WD has reached the Chandrasekhar mass.
The SDS may occur when a WD accretes matter via RLOF 
from any hydrogen-rich companion (e.g.,  main sequence or evolved star).
If the WD accumulates enough hydrogen on its surface such that steady 
burning can occur, the WD can increase in mass up to 1.4 M$_{\odot}$,  
carbon ignites explosively in the WD centre and the result is a SN Ia.
In the AM CVn scenario, we assume the result is a SN Ia if the CO WD
reaches 1.4 M$_{\odot}$ via stable RLOF from a helium-rich companion.  
We allow for different types of helium-rich donors in the AM CVn
scenario: helium stars (non-degenerate stars burning helium in the 
core or in a shell which have been stripped of their outer hydrogen 
envelope), helium white dwarfs, and hybrid white dwarfs (CO-rich core,
helium-rich mantle). 

We adopt two contrasting star formation rates (SFR): an instantaneous burst
at $t=0$ (elliptical galaxy) and a constant SFR for 10 Gyr (spiral galaxy).  
Both populations are then evolved up to 15 Gyr. 
The mass formed in stars in both cases is the same: $6 \times 10^{10}$
M$_{\odot}$, which corresponds to the stellar mass in the Milky Way \citep[MW;][]{KZS02}.
In each population we adopt a binary fraction of $50\%$ (2/3 stars in
binaries), though this fraction may be overestimating the binary population among 
low-mass stars \citep{Lad06}, and underestimating the binary fraction
among massive stars \citep{KF07}.  
All stars are evolved with solar-like metallicity ($Z=0.02$).
While the initial distributions representative of the physical
characteristics of ZAMS binaries, and the correct way in which to treat  
common envelope evolution and magnetic braking 
are all somewhat uncertain, our choices for various
distribution functions are constrained by available observations (see
below).  
The magnetic braking prescription which we adopt  
is that of \citet{IT03}, which is based on a two-component coronal 
model \citep[see Section 3.2 of][]{Bel08}.  

ZAMS masses ($M_{\rm ZAMS}$) span the entire mass range: $0.08-150
\msun$.  Single stars and binary primaries are drawn from a 3-component broken power
law initial mass function \citep{KTG93}, and secondary masses are obtained
from a flat mass ratio distribution (\citet{Maz92}, $q=$secondary/primary), which is
the canonical choice among population synthesis studies \citep{LYH06}.  
However, given the observational selection effects, the true mass ratio
distribution among ZAMS binaries remains unknown, though it is likely 
dependent upon stellar mass \citep[][]{Tri90,DM91}.  
It has been suggested that 
the mass ratio distribution among local spectroscopic field binaries
as well as young early-type stars is peaked near unity \citep{FSS05,KF07}.  
Initial orbital 
separations in our calculations span a wide range up to $10^{5}$ R$_{\odot}$ and are 
drawn from a distribution $\propto 1/a$ \citep{Abt83}, which has 
been found to be representative of the local population of {\em Hipparcos} 
binaries  \citep{LB07}.  Initial 
eccentricities are drawn from a distribution $\Psi(e)=2e$ \citep{DM91}.

Close binaries (and potential SN Ia progenitors) are believed to
encounter a phase of common envelope evolution, and this remains one of
the most poorly-understood phases in stellar astrophysics.  For this
reason, we present the delay times and rates of SNe Ia for two 
different CE parameterizations.  Some
comparison between different prescriptions for CE evolution have
been tested against observations for local double white dwarfs
\citep{NT05}.
It is unclear at this point how the common envelope phase should best
be treated in population synthesis studies, and currently detailed
models are not sophisticated enough to explore the parameter space in
detail \citep{RT08}.
For this work, we choose the `$\alpha$' prescription for CE
evolution \citep{Web84}, in which the orbital energy of the binary
is used to unbind the common envelope from the system.  
We choose two different values for the parameterization of common
envelope removal efficiency $\alpha_{\rm CE}$.  For our standard
model (Model 1) we choose $\alpha_{\rm CE} \times \lambda = 1$, where $\lambda$
  is a function of the donor envelope structure, and is of order unity
  \citep[see also][]{vVP06}.  As an alternative, we additionally include a model
with decreased common envelope removal efficiency in which
$\alpha_{\rm CE} \times \lambda = 0.5$ (Model 2).

\section{Delay Times}

{\em Model 1: $\alpha_{\rm CE} \times \lambda = 1$.}
We use the elliptical model, with all stars born at $t=0$ to demonstrate the delay
time distribution for the various progenitors. Figure 1 (top panel) shows the characteristic
delay times for Model 1, and the average and median delay times of all three progenitors
are also indicated.  The sharp cut-off near 15 Gyr is artificial, as evolution was 
only allowed to proceed for 15 Gyr. 

The DDS events follow a power-law like DTD, with a median of
$t_{\rm Med}=0.93$ Gyr and approximated functional form of $f(t) \propto 100
t^{-1}$. Only a small fraction $\sim 5\%$ are `prompt' \citep{MDP06} events with
delay times $t < 100$ Myr. The DDS events are expected to be found 
long after the star formation has ceased ($t \sim 10-15$ Gyr) therefore
we expect these progenitors to produce SNe Ia both in young (spiral and
starburst) and old (elliptical and bulges) host populations.
Even though we have allowed DDS progenitors to arise from mergers between any 
combination of CO and/or He WDs, all of our DDS systems originate from mergers of
CO-CO WD binaries, since mergers of He-He or CO-He WD binaries never exceed 
$1.4$ M$_{\odot}$.  
The evolution leading to the formation of a CO-CO WD binary usually starts with
two intermediate-mass stars (M$_{\rm ZAMS} \sim 3-9$ M$_{\odot}$) that evolve through a 
series of close interactions, the
first one typically being stable RLOF and the last one being a CE phase. Once a 
CO-CO WD is formed, the dominant mechanism for angular momentum loss is gravitational 
radiation, the merger timescale $\tau${\scriptsize$_{GR}$} $\propto 
a^{4}$ \citep{Pet64}.  
Given the initial distribution of orbital separations $\propto a^{-1}$,
and the evolutionary orbital change that is to first order the same for all 
DDS progenitors (reduction of orbital size by a factor of $\sim 10-100$ during the CE
phase), the delay time $t$ should likely then follow from the product of the 
initial distribution and the change in separation which occurs due to 
gravitational radiation: \\
$a^{-1} d\,a/d\,t \propto (t^{1/4})^{-1} t^{-3/4} 
\propto t^{-1}$. 
Thus, it is not surprising that the DTD follows a power law as
presented in Figure~1.

It is found that relatively speaking, the shortest delay times originate from the AM CVn
channel. The distribution is somewhat narrow, with the majority of events occurring
with $t < 2$ Gyr, with a median of  $t_{\rm Med}=0.59$ Gyr.
Most AM CVn progenitors originate from intermediate-mass stars (primaries M$_{\rm ZAMS} \sim 5-7$ M$_{\odot}$, secondaries M$_{\rm ZAMS} \sim 2-4$ M$_{\odot}$), however, in
contrast to DDS progenitors, they undergo two CE events. After two CE phases
the orbits of pre-AM CVn binaries are very close, which allows for rapid
orbital decay due to emission of GR, and the final RLOF starts with no
significant delay.  At the onset of the final RLOF (start of AM CVn phase), 
the accretor is a massive CO WD and the companion is either 
a helium star ($65$ \%) or a helium/hybrid WD ($35$ \%). 
Once the RLOF commences, it proceeds on a short-timescale with high mass
transfer rates ($\sim 10^{-5}-10^{-6}$ M$_{\odot}$ yr$^{-1}$), enabling the CO WD 
to rapidly build up to the Chandrasekhar mass and subsequently produce
a SN Ia explosion. 
Note that the potential SN Ia progenitor systems discussed here in the
framework of the AM CVn scenario are by no means a representation of the observed 
sample of $\sim 22$ AM CVns. The rare systems discussed here are more massive, 
ultra-compact systems (median $P_{\rm orb} \sim 10$ min) and thus have higher mass 
transfer rates than the population of AM CVns which is presently 
observed (median $P_{\rm orb} \sim 35$ min).  The typical, low-mass 
AM CVn binaries are long-lived systems which do not disappear from 
the observational population due to SN Ia disruptions.  There is a
clear observational bias against those AM CVn binaries which can potentially
produce a SN Ia.  

The SDS channel displays a rather flat (though somewhat decreasing) 
DTD over a wide
range $t \sim 1-15$ Gyr, with a median of $t_{\rm Med}=3.23$ Gyr.
The SDS systems are found to be binaries with a CO WD accreting from an evolved
star ($\sim 95\%$; mostly red giants) or a main sequence star ($\sim 5\%$).
Since the evolution leading to the formation of a CO WD is rather fast, the
delay time for the SDS is set by the evolutionary timescale of the evolved donor. 
That in turn is a strong function of initial ZAMS donor mass ($0.7 < $M$_{\rm ZAMS} 
< 2.7$ M$_{\odot}$; lifetimes $0.5-30$ Gyr). The longest delay times follow from  
progenitors whose donors, on average, had the smallest initial masses. Mass 
transfer rates can be initially high; up to $\sim 10^{-3}$ M$_{\odot}$ 
yr$^{-1}$, in the case of giants with $M_{\rm ZAMS} > 1$ M$_{\odot}$, though 
hydrogen accumulation is only fully efficient for a narrow range of 
mass transfer rates \citep[see][Figure 4]{Nom07}. Despite the high mass
transfer rate from the donor, only a fraction of the hydrogen accumulates on 
the white dwarf (accumulation rates of $\sim 10^{-7}$ M$_{\odot}$ yr$^{-1}$).  
For SDS progenitors with main sequence donors, the two stars are brought into 
contact due to loss of orbital angular momentum from magnetic braking 
(convective secondary) and to a lesser extent, gravitational wave emission.  
Mass transfer and accretion rates are much lower ($\sim 10^{-11}$ M$_{\odot}$ 
yr$^{-1}$), and the WD takes $\gtrsim 1$ Gyr to accrete up to the Chandrasekhar 
mass.  

{\em Model 2: $\alpha_{\rm CE} \times \lambda = 0.5$.}
The delay time distribution for Model 2 is shown in Figure 1 (bottom panel).  The change in the
treatment of the CE evolution leads to smaller orbital separations
after the common envelope phase, which affects the subsequent binary evolution
and thus leaves an imprint on the resulting DTD. 

As with Model 1, Model 2 DDS
progenitors involve only CO-CO WD binaries.  However with the
decreased CE efficiency, many potential DDS progenitors merge before
a detached white dwarf binary has been produced, thus there is a lack 
of DDS progenitors and there is an
increase in the number of merging WD + AGB core systems.  
The same power-law curve from the top panel of Figure 1 is shown 
for comparison.  The power-law-like shape of the DDS is present in Model 2 
for delay times $\lesssim 6$ Gyr with an additional pile-up of progenitors 
at short delay times ($t \lesssim 1 $ Gyr). The pile-up is 
due to the fact that in Model 2, DDS progenitor systems which survive the CE 
phase (e.g., binaries which do not merge in the common envelope) are 
on closer orbits upon emerging from the CE, thus they merge with
relatively shorter delay times ($\tau${\scriptsize$_{GR}$} $\propto 
a^{4}$).  
This `shift' to earlier delay times in the DTD from Model 1 to Model 2
serves to build a relatively stronger peak at short delay times, 
while at the same time decreases the number of DDS progenitors with 
delay times $\gtrsim 1$ Gyr.  This is the reason for the shorter 
median delay time in Model 2: $t_{\rm Med}=0.52$, vs. that of Model 1 
($t_{\rm Med}=0.93$).

The SDS channel of Model 2 is more efficient than that of Model 1 
since the WD and the 
stellar companion are brought on a closer orbit during the CE phase.  For this
reason, wider systems which would have evolved to become double white
dwarfs (e.g., AM CVn double degenerates) in Model 1 evolve 
into binaries with WDs accreting from non-degenerate companions in Model 2.  
In Model 2, the contribution of main sequence donors in the SDS channel is 
slightly higher since after the common envelope and formation of a CO WD,
in many cases the stars are close enough for the secondary to fill its Roche
lobe before evolving off of main sequence. 
The number of SDS progenitors
involving main sequence donors is doubled relative to Model 1 and now 
constitutes $10$ \% of the total SDS population ($90$ \% are giant or sub-giant
companions).  
SDS progenitors originate from systems with donor
masses $\sim 0.7 <$ M$_{\rm ZAMS} < 2.7$ M$_{\odot}$, many of 
which are M$_{\rm ZAMS} > 1.25$ M$_{\odot}$ thus have main sequence
lifetimes shorter than $\sim 5$ Gyr.  
The increased number of SDS progenitors at earlier delay times
leads to a decrease in the SDS median delay time from $\sim 3.2$
Gyr (Model 1) to $\sim 2.1$ (Model 2).  


The DTD of the AM CVn progenitor population of Model 2 is
notably different from that of Model 1, in that it is bimodal.  
The `fast' channel at delay times $\lesssim 3$ Gyr is still present,
and originates from progenitors with donor masses M$_{\rm ZAMS} \gtrsim 2.5$
M$_{\odot}$ which undergo two CE events.  
However in Model 2, the decreased orbital separation of post-CE
binaries allows for the formation of a new SN Ia progenitor channel, in 
which a CO WD initially accretes hydrogen from a low-mass main sequence 
star for several Gyr, as the main sequence star continues to fuse 
hydrogen into helium in its core \citep[see also][]{Pod08}.    
The main sequence star, which has been losing its outer layers in RLOF, 
has built up a significant helium core by the time its mass is
depleted to the hydrogen burning mass limit (0.08 M$_{\odot}$).  
Upon reaching the hydrogen-burning mass limit, the donor is said to be
degenerate, and since it is helium-rich (exposed core) we treat it as a helium
white dwarf.  There is a brief period (several Myr) where RLOF 
ceases due to the decreased size of the newly degenerate helium WD before 
the stars are brought in contact again due to emission of 
gravitational radiation.  Stable RLOF begins between the CO
WD and helium WD, until the CO WD accretes up to the Chandrasekhar
mass, exploding in a SN Ia.  This additional (and scarcely-populated) 
`slow' AM CVn channel
with delay times $> 6$ Gyr originates from progenitors with donor 
masses $0.7 < $ M$_{\rm ZAMS} < 1.0$ M$_{\odot}$, which only undergo 
one CE event, and evolve through an active cataclysmic variable (CV) phase for
$\gtrsim$ a few Gyr.  Under the Model 1 CE evolution, such a system 
would be found as a detached COWD + MS binary, and would not make a
CV, nor a SN Ia, in a Hubble time.    

\section{Rates}

Assuming a binary fraction of 50 \%, as we have done in this study, $0.17$ 
\% and $0.09$ \% of stellar systems (where a `system' represents either a single 
star or a binary) evolve into SNe Ia in Model 1 and Model 2, respectively.  
For an assumed binary fraction of 100 \%, $0.34$ \% and $0.18$ \% of binaries 
evolve into SNe Ia in Model 1 and Model 2, respectively.
Note that this applies to both elliptical and spiral populations as they
differ only in SFR but not in any evolutionary parameters other than
$\alpha_{\rm CE} \times \lambda$.  The above fractions translate into  
integrated efficiencies of 1 SN Ia per 2500 M$_{\odot}$ and 4700
M$_{\odot}$ of formed stars for Models 1 and 2, respectively, when 
a binary fraction of 50 \% is assumed.  For an assumed binary fraction
of 100 \% the integrated efficiencies are 1 SN Ia per 1500 M$_{\odot}$
and 1 SN Ia per 2800 M$_{\odot}$ for Models 1 and 2, respectively.  
Though the efficiency scales with the adopted binary fraction, the 
rates as a function of time, obviously, are different depending on 
the adopted SFR.   

A general summary of Model 1 and Model 2 rates is presented in Tables 
1 \& 2 where we show the   
rate of SNe Ia in SNuM, where 1 SNuM $\equiv 1$ SN ($100$ 
yr)$^{-1}$ ($10^{10}$ M$_{\odot}$)$^{-1}$ \citep{Man05}. 
We show the number of SNe Ia in SNuM for the 
three Chandrasekhar-mass models investigated in this work 
at 4 different epochs: $0.5$, $3$, $5$ and $10$ Gyr. 
SNuM rates are shown for our four galaxy models:
elliptical, Model 1; spiral, Model 1;  
elliptical, Model 2; spiral, Model 2.  

{\em Model 1}. 
In Figure 2 (top panel), SN Ia rates (number of SNe per unit time) are shown for 
our Model 1 elliptical galaxy ($\alpha_{\rm CE} \times \lambda = 1$, 
instantaneous starburst at $t=0$). The three progenitor types: DDS,
SDS and AM CVn are shown separately. 
The DDS rate declines with time, but DDS progenitors are found both at 
early and late times. The DDS events dominate over the other progenitor types. 
This dominance is very strong ($\sim 1-2$ orders of magnitude) and holds for the 
entire evolution of an elliptical galaxy ($0-15$ Gyr). The rate of SNe Ia from 
AM CVn systems is high for times $\lesssim 2$ Gyr and then rapidly declines.  
Since for a typical AM CVn progenitor (i) the CO WD forms early ($t_{\rm evol} < 100$ Myr)
and with a high mass ($M_{\rm COWD} \sim 1.1$ M$_{\odot}$; from primaries of initial mass 
$M_{\rm ZAMS} \sim 5-7$ M${\odot}$), and (ii) the orbital separation after two common envelopes 
is small\footnote{$a \sim 0.3 - 1 $R$_{\odot}$, where the primary is a CO WD and the companion is 
either a helium star ($M_{\rm He} \sim 0.4$ M$_{\odot}$) or a helium WD ($M_{\rm HeWD} \sim 0.3$ M$_{\odot}$).}, the two stars are brought into contact either by GR (He WD) or a combination of 
GR and evolutionary expansion (helium star).  Once the donor fills its Roche Lobe, accumulation is fully efficient for binaries with helium star donors (mass transfer rates $\sim 10^{-8}$ M$_{\odot}$ yr$^{-1}$), though in the case of helium WD donors, the mass transfer rate is initially higher ($10^{-5} - 10^{-6}$ M$_{\odot}$ yr$^{-1}$) due to the smaller stellar separation of these 
systems upon reaching contact ($\sim 0.3$ vs. $1.0$ R$_{\odot}$).  
For these relatively high mass transfer rates, only a fraction ($\sim 50$ \%) of the mass is accreted by the CO WD.  In either case, the 
mass accretion rate averaged over time from contact to SN Ia is found to be on the order of $10^{-7} - 10^{-8}$ 
M$_{\odot}$ yr$^{-1}$ for AM CVn SN Ia progenitors, thus it takes $\sim 10 - 100$ Myr for the CO 
WD to accrete up to the Chandrasekhar mass.  Therefore, the delay time is: $t_{\rm evol}$ $+$ ($\sim 10 - 100$) Myr ($+$ $t_{\rm GR}$ in the case of helium WD donors)$ = 0.2 - 2$ Gyr for a typical AM CVn SN Ia progenitor to produce a SN Ia.

The SDS rate maintains a nearly constant SN Ia rate through about $6 - 7$ Gyr and then drops by almost an order of magnitude. The increased rate at shorter 
delay times is due mainly to systems with evolved donors which encounter, on 
average, RLOF on shorter (evolutionary) timescales than their main sequence 
counter-parts for which the RLOF is encountered on longer (magnetic braking) 
timescales.  

The resulting rate of potential DDS SNe Ia varies substantially with time. At early 
times ($t \lesssim 1$ Gyr) the rates are very high $\sim 0.01$ yr$^{-1}$,
and then they gradually decrease to reach $\sim 0.0003$ yr$^{-1}$ at late
times ($t \gtrsim 10$ Gyr). The observed rates for elliptical galaxies are
estimated at the level of ${\cal R}_{\rm obs} \sim 0.0018 \pm 0.0006$ yr$^{-1}$
per unit ($10^{10} L^B_\odot$) of blue luminosity \citep{CET99}.   
As the blue luminosity of elliptical galaxies declines with time (after an early 
star formation episode), the rates presented in the top panel of Figure 2 should be corrected 
downwards at early times, while at later times they should be
increased if our rates are to be compared with those of typical ellipticals.  
Obviously, the burst of star formation on the order of $6 \times 10^{10} \msun$ 
would produce a blue luminosity larger than $10^{10} L^B_\odot$, while 10-15 Gyr
after the episode when stars more massive than $\sim 1 \msun$ have formed remnants 
and are no longer contributing to the galaxy's light, the blue luminosity is smaller 
than the normalising value.  
Since we do not really know the distribution of
age of the galaxies in the observed sample of ellipticals that were used in the SN
Ia rate estimate, we do not attempt to correct our synthetic rates for the evolution
of blue luminosity and we do not compare them directly to the observed
rates of \citet{CET99}.  However, we note that the observed rate is
consistent with our 
predicted rates for the DDS progenitor, while the predicted rates for other
progenitors (SDS and AM CVn) seem to be significantly too low. 

In Figure 2 (bottom panel), we show the SN Ia rates for the spiral galaxy model of Model 1. It is found 
that DDS rates of SNe Ia at the current epoch are $0.002$ yr$^{-1}$. At
first, the DDS rate increases with time (after the onset of star formation), 
then remains approximately constant until the star
formation stops leading to an overall decline in the rate. This behavior 
reflects the specific shape of the delay time distribution for the 
DDS combined with the SFR for our spiral galaxy model. 
The rates for SDS and AM CVn progenitors are much smaller and at the level of 
$\sim 10^{-4}$ yr$^{-1}$. SDS progenitors can generate SNe Ia long after
star formation has ceased (long delay times), while AM CVn events disappear 
shortly after the star formation has stopped (short delay times). 
For comparison, over-plotted are empirical rates of SNe Ia. The rates were 
adopted from \citet{CET99} for a Milky Way type spiral 
(Sbc-Sd) with a blue luminosity of $2 \times 10^{10}$ L$_{\odot}$, and the
rates are ${\cal R}_{\rm obs}=0.004 \pm 0.002$ SN Ia yr$^{-1}$. 
The DDS rate alone is consistent with the empirical rate of SNe Ia. The SDS 
and AM CVn SN Ia rates do not even come close to the empirical rate, and their 
addition to the DDS rate does not significantly affect the overall rates at 
any epoch.  
We note that our mass normalization which implies a constant star formation 
history for 10 Gyr results in a SFR at the level of $6 \mpy$. The global SFR 
in the MW may be somewhat lower: $\sim 3.5
\mpy$ \citep{Cox00,Osh08}, and in that case the DDS rates are 
only marginally consistent with the observed \citet{CET99} rates.  
On the other hand, it
has been suggested that the SFR of the MW has been decreasing with time, 
only reaching $\sim 3.5 \mpy$ at the current epoch \citep[][see sect. 2.2]{NYP01,NYP04}.
If such an estimate had been used the average SFR of the MW is found
at the level of $\sim 8 \mpy$, and our results would scale up, being consistent with 
the DDS scenario as the major SN Ia contributor in MW-like spiral galaxies,
as long as the \citet{CET99} rates are being used for comparison.

{\em Model 2}. 
There is a marked decrease, by nearly a factor of 2, in the total number of SNe Ia progenitors 
in our model with decreased CE removal efficiency.  The overall decrease is due to the fact 
that the most dominant potential channel, the DDS, is only $\sim 50$ \% as efficient, since 
a larger fraction of binaries will merge in the common envelope phase 
rather than surviving the CE to subsequently form a double white dwarf.  

In Figure 3 (top panel), SN Ia rates are shown for our Model 2 elliptical galaxy 
($\alpha_{\rm CE} \times \lambda = 0.5$, instantaneous starburst at $t=0$). 
DDS SNe Ia progenitors continue to outnumber the SDS and AM CVn
progenitors, however there are some notable differences.   
For short delay times $t \sim 1 $ Gyr, the DDS
rates are nearly a factor of 2 lower than they are for Model 1.  
Then at later times, the Model 2 DDS rates are a factor 
of $\sim 3$ below those of Model 1, reaching $\sim 9 \times 10^{-5}$
yr$^{-1}$ at delay times of 10 Gyr (vs. $3 \times 10^{-4}$ yr$^{-1}$ for Model 1).     
Despite the lower DDS rates of Model 2, 
potential progenitors are found at all delay times, as
they are in Model 1.  The CE efficiency of Model 1 
allows for DDS progenitors to be drawn from a wider range (going to
smaller values) among the distribution of initial separations, where
as progenitors with small initial separation in Model 2 are
removed from the DDS population in mergers during the CE phase.  
The Model 2 SDS channel is more efficient (by a factor of 3) than the
Model 1 SDS channel since post-CE WD + MS binaries are found on closer
orbits.  One major difference
between the elliptical galaxy Ia rates of Model 1 and Model 2 
is that the SDS rates match those of the DDS rates for delay times $\sim
2.5 - 5.5$ Gyr ($\sim 0.0002$ yr$^{-1}$).  
Since the majority of the donors are evolved 
stars (giants or sub-giants), the delay involves two components: the main sequence lifetime 
of the donor (a few Gyr for a donor to become a giant) and the accretion 
timescale, over which the primary WD can increase its mass to the Chandrasekhar 
mass (10-100 Myr).  Thus the main sequence lifetime of the donor is what 
sets the delay times for the SDS DTD.  
Since the stars are found on 
closer orbits after the common envelope, RLOF is encountered between
the WD and the non-degenerate companion more often in Model 2 
(typically when the donor is a sub-giant).  
Rates of potential AM CVn SN Ia are lower than those of Model 1 for
elliptical galaxies and are at the level of $\sim 0.0002$ yr$^{-1}$. 
For the majority of the progenitors the delay times are very short, so as in
Model 1, these type of events are expected only in young host galaxies or
in regions with ongoing star formation. `Fast' AM CVn progenitors are more rare in 
Model 2 since these systems more readily merge in one of the two common envelope 
phases that lead to the formation of these progenitors. 
In contrast to Model 1, there is a small contribution of the `slow' AM CVn 
progenitors (long delay times) in Model 2.
The AM CVn channel is outnumbered by both the SDS and DDS channels at all epochs 
in the Model 2 elliptical galaxy.  

In Figure 3 (bottom panel), SN Ia rates are shown for our Model 2 spiral galaxy.   
It is found that DDS rate of SNe Ia at the current epoch ($10 $ Gyr) is
$0.001$ yr$^{-1}$; a factor of two below that of the Model 1 spiral galaxy. 
At first, the DDS rate increases and then remains
fairly constant until star formation ceases at 10 Gyr.  At all epochs
(particularly during star formation), 
the DDS channel rates significantly outnumber the SDS and AM CVn channels, but to a
lesser degree than when compared with Model 1.  The SDS channel
also exhibits a relatively constant rate at times later than $\sim
1$ Gyr.  At the current epoch the SDS rates are at the level of 
$ 0.0002$ yr$^{-1}$; a factor of $\sim 5$ below the DDS rates.  The
rates arising from AM CVn progenitors are fairly negligible at all
epochs ($< 10^{-4}$ yr$^{-1}$).  Even when the rates from all three progenitor
channels in the considered galaxy model are combined (0.0012 SNe Ia yr$^{-1}$), 
the SN Ia rate at the current epoch falls below the empirically-derived rate from 
\citet{CET99} by roughly a factor of two.  

It is worth comparing our model galaxy rates to the rates presented in 
\citet{Man05}.  In that study, \citet{Man05} derive SN rates 
for galaxies of various morphological types, and present the SN rates in SNuM 
(as inferred from $K$-band luminosity measurements).  We cannot directly compare 
our rates in Tables 1 \& 2 to those of \citet{Man05} since we do not know 
the exact ages of the galaxies in their sample.  However, we note that our 
Model 1 elliptical galaxy SN Ia rate at 10 Gyr\footnote{A typical age for local ellipticals; see Mannucci et al. 2005 Section 7.} is $0.005$ SNuM (see Table 1 DDS rates), which is nearly a 
factor of 10 below the SN Ia rate in E/S0 galaxies presented by \citet{Man05}: 
$0.044^{+0.016}_{-0.014}$ SNuM (see their Table 2). 
We note as well that for the same model galaxy, at $t = 500$ Myr (shortly after a burst of star 
formation at $t=0$) we obtain a SN Ia rate of $\sim 0.18$ SNuM (mostly 
via the DDS channel with some contribution from AM CVn), 
which is about a factor of $2$ lower than the range of SN Ia rates found for star 
forming Irregular galaxies in \citet{Man05} ($0.77^{+0.42}_{-0.31}$).  
The \citet{Man05} SN Ia rate for S0a/b spirals is found to be 
$0.065^{+0.027}_{-0.025}$ SNuM, which matches our Model 1 
spiral galaxy rates at 5 Gyr ($0.065$ SNuM; mostly DDS \& AM CVn progenitors).  The SN Ia rate 
of Sbc/d spirals from \citet{Man05} is $0.17 ^{+0.068}_{-0.063}$, and matches 
our Model 1 spiral SN Ia rate only at very early times ($\sim 0.5 - 1$ Gyr, see Table 1).  

For the Model 2 elliptical galaxy, we find a low SN Ia rate of $\sim 0.001$ SNuM 
at 10 Gyr, which is over an order of magnitude below the rate for E/S0 galaxies in \citet{Man05} 
($0.044^{+0.016}_{-0.014}$ SNuM).  
The rate at 500 Myr for the same galaxy is $\sim 0.14$ SNuM (see Table 2); mostly arising from  
DDS progenitors with some contribution from AM CVn and SDS.  This rate is a factor of a few below the Ia 
rates for Irregular galaxies in \citet{Man05} ($0.77^{+0.42}_{-0.31}$).  In comparing our 
Model 2 spiral rates, we find that our SN Ia rate at 3 Gyr ($\sim 0.07$ SNuM; mostly DDS progenitors with some contribution from SDS) is within the range 
of rates presented in \citet{Man05} for S0a/b spirals ($0.065^{+0.027}_{-0.025}$), while only 
our spiral rates for Model 2 at very early times ($< 1$ Gyr; $0.13$ SNuM, DDS) are high enough 
to match those of \citet{Man05} for Sbc/d type galaxies ($0.17^{+0.068}_{-0.063}$).  
We note that in general, our rates (per unit mass) are lower than those of \citet{Man05},  
indicating that perhaps other channels leading to the formation of SNe Ia should 
be considered in evolutionary studies (e.g., single stars, sub-Chandrasekhar mass SNe Ia). 
We also note however that our predicted rates as a function of time (at least for the 
DDS) are consistent with the observed rates presented by \citet{CET99}.

We find that in general the DDS outnumber SDS and AM CVn progenitors.  
This effect is somewhat more pronounced in Model 1 (a factor of $\gtrsim
10$) than in Model 2 (a factor of $\sim 5$), but the reason why is clear for both
models.  
The occurrence rate of a CO-CO WD merger with a total mass $\geq 1.4$ M$_{\odot}$
(DDS) is higher than that of building up a CO WD's mass to $\sim 1.4 \msun$ 
via stable mass transfer (SDS) in a binary. 
Formation efficiencies in both of the above cases are very low, after all SNe Ia 
are rather rare events. However,  
relatively speaking it is easier to find a pair of two CO WDs (DDS), each with 
a mass of $\gtrsim  0.7 \msun$ which is a typical mass for CO WDs, than it is for 
a CO WD to double its mass through accretion (SDS/AM CVn). 
This finding is a consequence of the recent updates on accumulation 
physics calculations \citep{HKN99,KH99,Nom07} that we have adopted in 
our evolutionary study. Basically, 
the accumulation onto a WD is hampered by a number of processes that tend to
remove matter which is transferred from the companion in a close binary (e.g., 
nova explosions, He-shell flashes, optically thick winds from the surface
of an accreting WD), in some cases leading to the disruption of
an accreting WD before it reaches the limiting Chandrasekhar mass; i.e.,
accretion of $\sim 0.1 \msun$ of a He-rich layer which ignites and
disrupts the underlying WD 
\citep[sub-Chandrasekhar mass SN Ia, see e.g.,][]{KH99}. Since, {\em (i)} there is 
a rather narrow range of mass transfer rates which may lead to efficient accumulation 
and {\em (ii)} there are not that many binary configurations (and we have considered 
the entire range for our adopted evolutionary model) that can sustain 
mass transfer for a 
prolonged period of time, the SDS and AM CVn channels are found to produce
SNe Ia at very low rates.

\section{Comparison With Other Studies}

The recent theoretical study of \citet{HKN08} 
finds a SDS delay time distribution which follows a power law. 
In their study, \citet{HKN08} incorporate a new mass stripping 
effect (based on \citet{HKN99}), where in the case of high 
mass transfer rates the WD blows an optically thick wind 
strong enough to `strip' material from a main sequence or 
giant donor.  This effect in return stabilizes mass transfer,
enabling the binary to avoid a CE phase even in the case of 
a relatively massive ($\sim 6$ M$_{\odot}$) donor.  
The result is 
that the WD can accrete stably up to the Chandrasekhar mass, with 
a wider range of potential progenitor donor ZAMS masses: 
$0.9 - 6$ M$_{\odot}$ in \citet{HKN08} vs. $0.7 - 2.7$ M$_{\odot}$ in 
our current study.  Even though we allow for SN Ia progenitors to form 
from any initial mass spanning the initial mass function, our 
SDS SNe Ia only derive from binaries involving low-mass donors since we do
not take into account this stripping effect.  
We note that the \citet{HKN08} model predicts the presence of a thick disc 
of hydrogen-rich circumstellar material around the SN Ia progenitor.  If 
such a circumstellar torus were present around the majority of progenitors 
of SNe Ia, one would likely expect to observe hydrogen in their spectra, 
though to date less than 1\% of SNe Ia have shown any signature of associated 
hydrogen \citep[see e.g.,][]{HP06}.  

\citet{HP04} investigated SNe Ia progenitors from WD + MS and WD + evolved 
binaries with very specific binary configurations using population synthesis.  
In that work, they do not present rates of SNe Ia derived from other possible 
formation channels of SNe Ia so we cannot compare DDS rates. 
The Galactic rate 
of SDS SNe Ia for their model which most closely matches our standard
(Model 1) parameters is $\sim 6 \times 10^{-4}$ yr$^{-1}$. This value is an order of 
magnitude above our SDS spiral galaxy Model 1 rate of $\sim 6 \times 10^{-5}$ yr$^{-1}$, 
though is still nearly an order of magnitude lower than the empirical
SN Ia Galactic rate of \citet[][$4 \times 10^{-3}$ yr$^{-1}$]{CET99}.  
The rate for the \citet{HP04} model which most closely matches 
Model 2 is $\sim 10^{-3}$ SNe Ia yr$^{-1}$, which is close 
(though still below) the Galactic rates of \citet{CET99}, and is a 
factor of 5 times higher than our Model 2 SDS rates 
($\sim 2 \times 10^{-4}$ yr$^{-1}$).  

It was pointed out by \citet{HP04} that 
their prescription for hydrogen accumulation is more efficient than that used by 
other authors \citep[i.e.,][]{YL98}. It is also more efficient than the prescription
we have adopted in this study.
The range of hydrogen accretion rates onto WDs which leads to stable
burning (see section 2)
and efficient mass accumulation is uncertain, and it is possible that stable 
hydrogen burning may occur for a wider range of accretion rates, in turn 
allowing for higher SNe Ia rates following from the SDS channel as allowed
in \citet{HP04}. However, comparison of model hydrogen-accreting WDs on the 
H-R diagram with supersoft X-ray sources \citep{Nom07} indicates that the 
prescription for stable hydrogen burning in a thin shell (and adopted here) is 
consistent with observations.  

It is worth noting that the delay time distributions of \citet{Gre05},
derived using analytical formulations, produce a 
DDS DTD shape which is similar to ours: peaked at short ($< 1$ Gyr) delay 
times, followed by a smooth drop-off as a function of time, due 
to the dependence of the delay time on the timescale associated with 
gravitational wave emission.  A similar trend is also found 
for the delay times of DDS progenitors in \citet[][Fig. 2]{YL00}.  
The \citet{Gre05} study also determined that the shape of the 
DTD arising from SDS progenitors was more flat than when compared 
to that of the DDS, and that the SDS delay time depended upon 
the main sequence lifetime of the secondary star (see their section 5),
which is consistent with our findings.    

Delay times of SNe Ia were calculated by \citet{BBR05}. It was
found that the merger of two white dwarfs was consistent with an 
empirical delay time estimate of $\sim 3$ Gyr 
\citep{Str04}\footnote{Delay times were computed by adopting a
    cosmic star formation history based on that of \citet{MDP98}.}, and that
WDs accreting from non-degenerate stars could potentially explain the
observed delay times if a low common envelope efficiency is used
($\alpha_{\rm CE} \times \lambda = 0.3$) and if it is presumed a priori that WD
mergers contribute negligibly or not at all to the SNe Ia
population. The above results were obtained with an earlier version of
the {\tt StarTrack} code; the code was recently updated to include the
most recent accumulation rates \citep[e.g.,][]{Nom07}.  All other
recent revisions relevant for low- and intermediate-mass binary
evolution in the code are described in detail in \citet{Bel08}.  
Additionally, in the \citet{BBR05} study, very different
criteria were adopted for SNe Ia; the majority of their supernovae
originated from sub-Chandrasekhar mass events and it was
permitted that the merger of two WDs of any type (including ONeMg WDs) with a
total mass exceeding 1.4 M$_{\odot}$ led to a SN Ia.  In the current
study we also note that the number of sub-Chandrasekhar mass events
exceed the number of Chandrasekhar mass progenitors.  The weakness of the
sub-Chandrasekhar model is that much of the outer (and fastest moving)
material is believed to burn to nickel with very few intermediate 
mass elements \citep{LA95}.  The resulting spectra do not match
current observations of normal supernovae, and the best fits are for
sub-luminous supernovae.  Only $\sim 6$ sub-luminous SNe Ia were
recently reported \citep{Kas08} as compared
with 36  \citep{Rie98} or 42 \citep{Str04} normal ones discovered only 
in the {\em Hubble} surveys.

\section{Discussion}

We have evolved single and binary stars using the population synthesis 
code {\tt StarTrack}, and have analyzed the resulting delay times (time 
from binary formation at $t=0$ to SN Ia) of potential Type Ia supernovae.
We have considered possible SNe Ia progenitors arising from three formation 
channels: Double Degenerate Scenario (white dwarf mergers), Single 
Degenerate Scenario (hydrogen-rich accretion on to a WD) and the AM Canum 
Venaticorum scenario (helium-rich accretion on to a WD). Additionally, we have 
computed SN Ia rates for two galaxy types: an elliptical galaxy with a 
starburst at $t=0$ and a spiral galaxy with a constant SFR, and in
each case we have tested the impact of the common envelope removal
efficiency for two different parameterizations: 
$\alpha_{\rm CE} \times \lambda =1$ (Model 1), and 
$\alpha_{\rm CE} \times \lambda =0.5$ (Model 2).  

Our SN Ia rates (century)$^{-1}$ ($10^{10}$M$_{\odot}$)$^{-1}$ have been presented in Tables 1 and 2.  
We reiterate that the rates which we have derived in this 
work are `local' (no redshift evolution) rates. It is still interesting to note that in 
calculating the volumetric SN Ia rate out to $z=0.12$ fitting the 
A $+$ B model, \citet{Dil08} found A $= (2.8 \pm 1.2) \times
10^{-14}$ SNe Ia yr$^{-1}$ M$_{\odot}^{-1}$, which is most similarly
matched by our DDS Model 1 rates for Milky Way-like spirals 
($\sim 3 \times 10^{-14}$ SNe Ia yr$^{-1}$ M$_{\odot}^{-1}$ at 10 Gyr). 
This 2-component or ``A$+$B'' model
\citep[see][]{MDP06,SB05}
assumes that the SN Ia rate is a function of stellar mass density and the
SFR, and allows for fitting prompt and tardy SN Ia 
populations.  We note that for Model 1, DDS progenitors dominate the overall SN Ia 
rates, whether it is for elliptical or spiral hosts. For
Model 2, the overall rates of SNe Ia are a factor of two lower than in
Model 1, though SNe Ia progenitors arising from the DDS and SDS
channels are found in equal numbers for delay times $\ 2.5 - 5.5$ Gyr.  

For all models considered, we expect both very short 
($\lesssim 1$ Gyr) and long ($\sim 10-15$ Gyr) SN Ia delay times 
if they originate from DDS progenitors.  
Could such a population explain the bimodal distribution of delay times derived
from some observations?  In principle, one may expect 
such a result; if DDS progenitors are dominant in both old and young
galaxies, the empirically-derived delay times may appear to be bimodal and
erroneously point toward two different progenitor populations. 
However, the reported bimodal DTD among SNe Ia 
\citep[][i.e., 50 \% of SNe Ia having `prompt' delay times $\lesssim
  100$ Myr]{MDP06} is not reproduced in any of our models.  
For our standard model, only $\sim 5\%$ of systems
have delay times below $100$ Myr, with $50$ \% of our SNe Ia occurring
within $\sim 800$ Myr since the starburst. 
The model which comes closest to reproducing such a bimodal delay 
time distribution is the DDS channel of Model 2, 
where there is a large fraction of SNe Ia
progenitors with short delay times (50\% with $t < 500 $ Myr).  
If in fact such a large fraction
of SNe Ia occur within 100 Myr of star formation, it may pose a very
interesting problem for binary evolution to explain.  Some
alternatives are already being considered, for example a single star
SN Ia progenitor \citep{Iben83,Mao08}.  
To properly approach this issue one needs to fold our evolutionary
calculations with the cosmic star formation history and distribution of
galaxy types and mass as a function of redshift.  
However, it was pointed out that the constraints on observational 
DTDs which incorporate convolution with 
assumed cosmic star formation history may not be as strong 
as they are claimed to be \citep{FPH06}.

For our standard (Model 1) set of evolutionary parameters, the DDS
dominates (roughly 90\%) the rate of Type Ia SNe in both spiral and
elliptical galaxies.  The rate of DDS SNe Ia is consistent with the
observed SN Ia rate of \citet{CET99}.  The DTD of potential DDS SNe Ia follows a smooth
power law distribution through a Hubble time.  
The SDS DTD is mostly 
flat throughout a Hubble time, while the AM CVn SNe contribute 
events mostly at short delay times, but neither 
contribute significantly to the total SN
Ia rate for spiral-like galaxies. 
The DDS DTD, following a power law, is consistent with the findings of
\citet{Tot08}, whose observationally-derived delay time of SNe Ia in
intermediate-redshift elliptical galaxies follows  
a featureless power law $\propto t^{-1}$ for $0.1 < t < 10$ Gyr.  

For Model 2, the DDS
dominates (83\%) the rate of Type Ia SNe in both spiral and
elliptical galaxies.  The DDS rate is below the observed
SN Ia rate of \citet{CET99} by a factor of two.  The DTD
of potential DDS SNe Ia does not exhibit as strong of a
power-law shape as Model 1, since relatively more white dwarf binaries merge  
at early delay times, and more would-be DDS progenitors merge during 
the CE phase and thus never produce a SN Ia.    
The SDS DTD is most prominent $\sim 2 - 6$ Gyr (at which times the
SDS rates match those of the DDS) and at later times the SDS contributes very little 
to the overall rates.    
The AM CVn SNe contribute some events mostly at short delay times, with the 
rates being at a very low level at long delay times.

As noted previously, all DDS SNe Ia progenitors are the result of a
merger of a CO-CO WD
binary. Mergers between CO-He WDs are relatively less common than He-He or
CO-CO WD mergers since often times mass transfer is stable, and upon reaching 
contact the CO-He binary will enter an AM CVn phase rather than coalesce.   
In any case, mergers of He-He or CO-He WDs are not massive enough to lead to 
SNe Ia, though they may lead to other very interesting phenomena, as they are 
likely precursors to helium-burning hot subdwarfs or R CrB stars
\citep{Web84}.  There is a marked decrease in the number of double
white dwarfs which are formed in Model 2 relative to Model 1 due to
a heightened number of mergers which are encountered during the CE
phase. 

Typically, population synthesis calculations
produce the right number of Chandrasekhar mass WD-WD mergers, yet usually produce an
order of magnitude too few SDS supernovae to explain the
observed rate estimates \citep[see][for a review]{Liv01}.
As it turns out, most hydrodynamical white dwarf merger calculations 
result in the less massive white dwarf being disrupted in
a few orbits, causing the more massive white dwarf to accrete at an
extremely high rate \citep{RS95,BTH89}.
At very high accretion rates, the white dwarf mergers are believed to collapse to
form a neutron star, not produce a SN Ia thermonuclear explosion
\citep{ML90,SN85,SN98,WW94}. Some new studies have found that some mergers might produce
Type Ia supernovae, but the merger conditions must be severely refined \citep{YPR07}.

Either population synthesis studies with the adopted accretion physics  
are missing active progenitor paths for Type Ia supernovae, or the physics 
of merger calculations are incorrect. All of our results and conclusions 
were based on one evolutionary model with a specific (albeit the most
updated) set of accumulation rates and have been presented from the standpoint 
of population synthesis 
(i.e., thus far we have ignored the information provided by merger calculations).  
\citet{YPR07} predicted that only DDS systems with small mass ratios 
$q<0.4$ can produce SN Ia explosions, while the rest will end up in accretion 
induced collapse and neutron star formation. If we had adopted this as an
additional criterion the implications would have been 
rather dramatic for our results.  At the time of the merger, only $\sim 0.2\%$ (0 \%) of the DDS systems
from Model 1 (Model 2) have $q<0.4$, since a low $q$ usually leads to
stable  mass transfer and not a merger.  
If the DDS rates had been decreased so drastically, none of our calculated SN Ia
progenitors would be able to match the observed rates. 

If we had relaxed our assumption on Chandrasekhar mass explosions and we had
incorporated sub-Chandrasekhar mass explosion models - 
ignition of a degenerate layer of He-rich material accumulated on the surface of 
a WD\footnote{The sub-Chandrasekhar mass model falls under 
our AM CVn formation channel.} 
 \citep{WW94,KH99} - the Galactic rates for AM CVn would increase from 
$\lesssim 10^{-4}$ yr$^{-1}$ to $\sim 10^{-3}$ yr$^{-1}$ 
for Model 1 and to $\sim 5 \times 10^{-4}$ for Model 2.  We note that in 
this estimate we have allowed sub-Chandrasekhar mass explosions to
occur only for white dwarfs with masses $M_{\rm wd} 
> 1 \msun$, since for lower WD masses the explosions 
would not look like those observed for typical Type 
Ia supernovae \citep[e.g., ][]{HN00}.  

If indeed white dwarf mergers (DDS) cannot lead to  
thermonuclear explosions, one needs to consider alternatives for increasing 
the rates of SDS and AM CVn progenitors, either by moving away from the standard
evolutionary model, or widening the range for efficient accumulation
rates onto white dwarfs.  

\acknowledgements 

AJR acknowledges a New Mexico State University Graduate Research 
Enhancement Grant and a Merit-based Enhancement Award, 
as well the {\em Supernova Rates 2008} LOC for support.  
KB acknowledges the support from KBN grant N N203 302835.
AJR and KB are thankful to L.Bildsten, K.Shen, R.Taam, 
L.Strolger, P.Podsiadlowski, K.Nomoto and T.Totani 
for informative discussion. AJR is additionally 
grateful to J.Grindlay for stimulating discussion, to J.Holtzman 
for useful suggestions while 
preparing this manuscript, and thanks B.Dilday and 
G.Nelemans for helpful information.

\clearpage

\begin{deluxetable}{lcc}
\tablewidth{400pt}
\tablecaption{Rates of SNe Ia Progenitors (SNuM) for Model 1}
\tablehead{Rate [($100$ yr)$^{-1}$ $10^{10}$M$_{\odot}^{-1}$] as a Function of Time.}
\startdata
DDS  & Elliptical & Spiral                \\  
\hspace*{0.5cm} 0.5 Gyr & $1.6 \times 10^{-1}$ & $2.0 \times 10^{-1}$ \\ 
\hspace*{0.5cm}  3 Gyr  & $2.3 \times 10^{-2}$ & $8.0 \times 10^{-2}$ \\
\hspace*{0.5cm}  5 Gyr  & $1.2 \times 10^{-2}$ & $6.0 \times 10^{-2}$ \\
\hspace*{0.5cm} 10 Gyr  & $\sim 5 \times 10^{-3}$ & $3.3 \times 10^{-2}$ \\
\hspace*{0.5cm} & & \\
SDS                    & &  \\
\hspace*{0.5cm} 0.5 Gyr & $\lesssim 10^{-3}$ & $ 0 $\\
\hspace*{0.5cm}  3 Gyr  & $\sim 3 \times 10^{-3}$ & $ \lesssim 10^{-3}$\\
\hspace*{0.5cm}  5 Gyr  & $\sim 1 \times 10^{-3}$ & $ \sim 10^{-3}$ \\
\hspace*{0.5cm} 10 Gyr  & $\lesssim 10^{-3}$ & $\sim 10^{-3}$\\
\hspace*{0.5cm} & & \\
AM CVn                &  & \\
\hspace*{0.5cm} 0.5 Gyr & $ 2.2 \times 10^{-2}$ & $ \sim 10^{-2} $ \\
\hspace*{0.5cm}  3 Gyr  & $ < 10^{-3}$ & $\sim 5 \times 10^{-3}$ \\
\hspace*{0.5cm}  5 Gyr  & $ \lesssim 10^{-4}$ & $\sim 4 \times 10^{-3}$  \\
\hspace*{0.5cm} 10 Gyr  & $ 0 $ & $\sim 1 \times 10^{-3}$ \\
\enddata
\label{tab1}
\end{deluxetable}

\begin{deluxetable}{lcc}
\tablewidth{400pt}
\tablecaption{Rates of SNe Ia Progenitors (SNuM) for Model 2}
\tablehead{Rate [($100$ yr)$^{-1}$ $10^{10}$M$_{\odot}^{-1}$] as a Function of Time.}
\startdata
DDS  & Elliptical & Spiral                \\  
\hspace*{0.5cm} 0.5 Gyr & $1.3 \times 10^{-1}$ & $1.3 \times 10^{-1}$ \\ 
\hspace*{0.5cm}  3 Gyr  & $\sim 5 \times 10^{-3}$ & $6.0 \times 10^{-2}$ \\
\hspace*{0.5cm}  5 Gyr  & $\sim 3 \times 10^{-3}$ & $3.5 \times 10^{-2}$ \\
\hspace*{0.5cm} 10 Gyr  & $\sim 10^{-3}$ & $1.4 \times 10^{-2}$ \\
\hspace*{0.5cm} & & \\
SDS                    & &  \\
\hspace*{0.5cm} 0.5 Gyr & $\sim 2 \times 10^{-3}$ & $ 0 $\\
\hspace*{0.5cm}  3 Gyr  & $\sim 5 \times 10^{-3}$ & $\sim 8 \times 10^{-3}$\\
\hspace*{0.5cm}  5 Gyr  & $\sim 3 \times 10^{-3}$ & $\sim 6 \times 10^{-3}$ \\
\hspace*{0.5cm} 10 Gyr  & $\sim 10^{-4}$ & $\sim 3 \times 10^{-3}$\\
\hspace*{0.5cm} & & \\
AM CVn                &  & \\
\hspace*{0.5cm} 0.5 Gyr & $\sim 5 \times 10^{-3}$ & $ < 10^{-3} $ \\
\hspace*{0.5cm}  3 Gyr  & $ 0 $ & $\sim 10^{-3}$ \\
\hspace*{0.5cm}  5 Gyr  & $ 0 $ & $\sim 10^{-3}$  \\
\hspace*{0.5cm} 10 Gyr  & $< 10^{-4} $ & $< 10^{-3}$ \\
\enddata
\label{tab2}
\end{deluxetable}

\clearpage

\begin{figure*}
\vspace{-1cm}
\includegraphics[width=\textwidth]{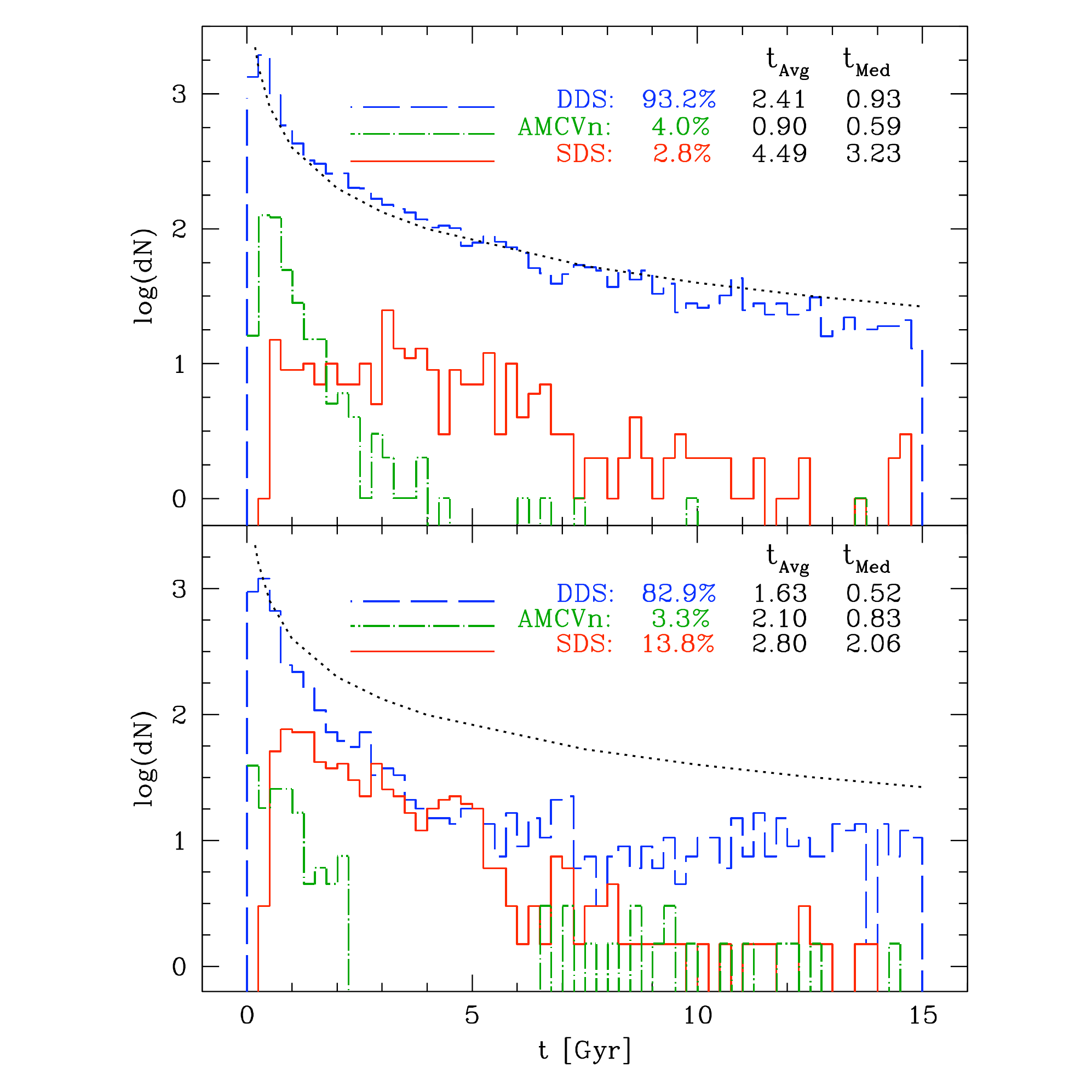}
\caption{Delay time distribution showing relative contributions from 
the three SN Ia formation channels considered in this work for our
elliptical galaxy (instantaneous burst of star formation at $t=0$): 
DDS (blue dashed), AM CVn (green dot-dash) and SDS (red solid).  Average and median 
delay times are indicated.  
Top panel: Model 1, $\alpha_{\rm CE} \times \lambda =                               
1.0$. The thin dotted line 
represents a power law function of form $f(t)=100 \, t^{-1}$ and follows the DDS 
delay time distribution reasonably well.  Bottom panel: Model 2, 
$\alpha_{\rm CE} \times \lambda = 0.5$.  The power law is shown for comparison.}
\label{plotone}
\end{figure*}

\clearpage

\begin{figure*}
\vspace{-1cm}
\includegraphics[width=\textwidth]{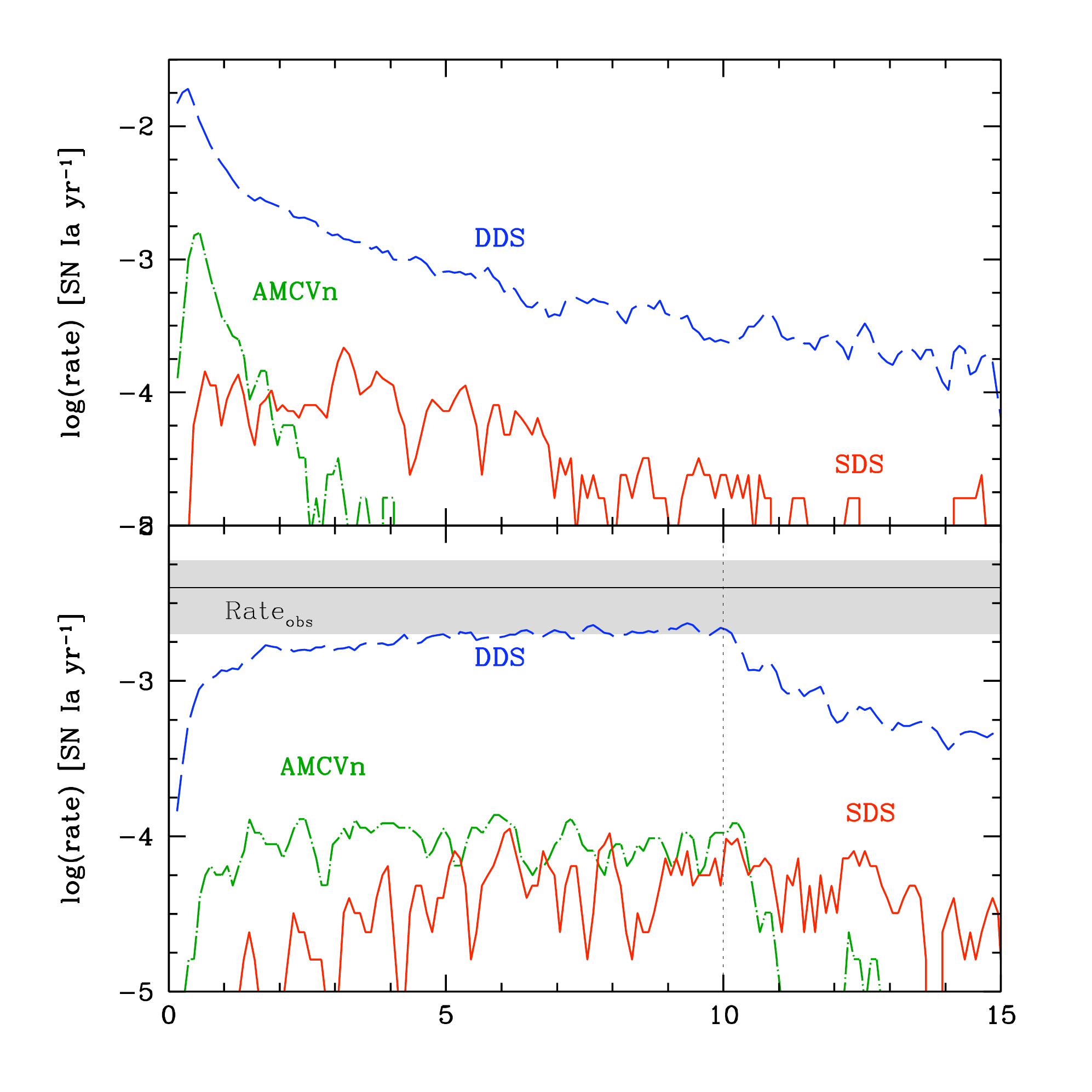}
\caption{Rates of Type Ia supernovae (number per year) for the Model 1 galaxies.  
The DDS (blue dashed), AM CVn (green dot-dash) and SDS (red solid) 
channel rates are shown. 
Top panel: elliptical galaxy whose total mass in formed stars is 
M$_{\rm tot} = 6 \times 10^{10}$ M$_{\odot}$ (delta function starburst at $t=0$).  
Bottom panel: spiral (constant star formation) galaxy whose total mass in formed 
stars at 10 Gyr is M$_{\rm tot} = 6 \times 10^{10}$ M$_{\odot}$. 
Approximating the star formation history of the MW
to be constant,
the current Galactic SN Ia rate (shown with vertical dotted line)
considering the DDS alone is 0.002 yr$^{-1}$.  This is consistent
with the empirical rate as indicated by the shaded region in the
plot \citep[$0.004 \pm 0.002$ yr$^{-1}$,][]{CET99}.}
\label{plottwo}
\end{figure*}

\clearpage

\begin{figure*}
\vspace{-1cm}
\includegraphics[width=\textwidth]{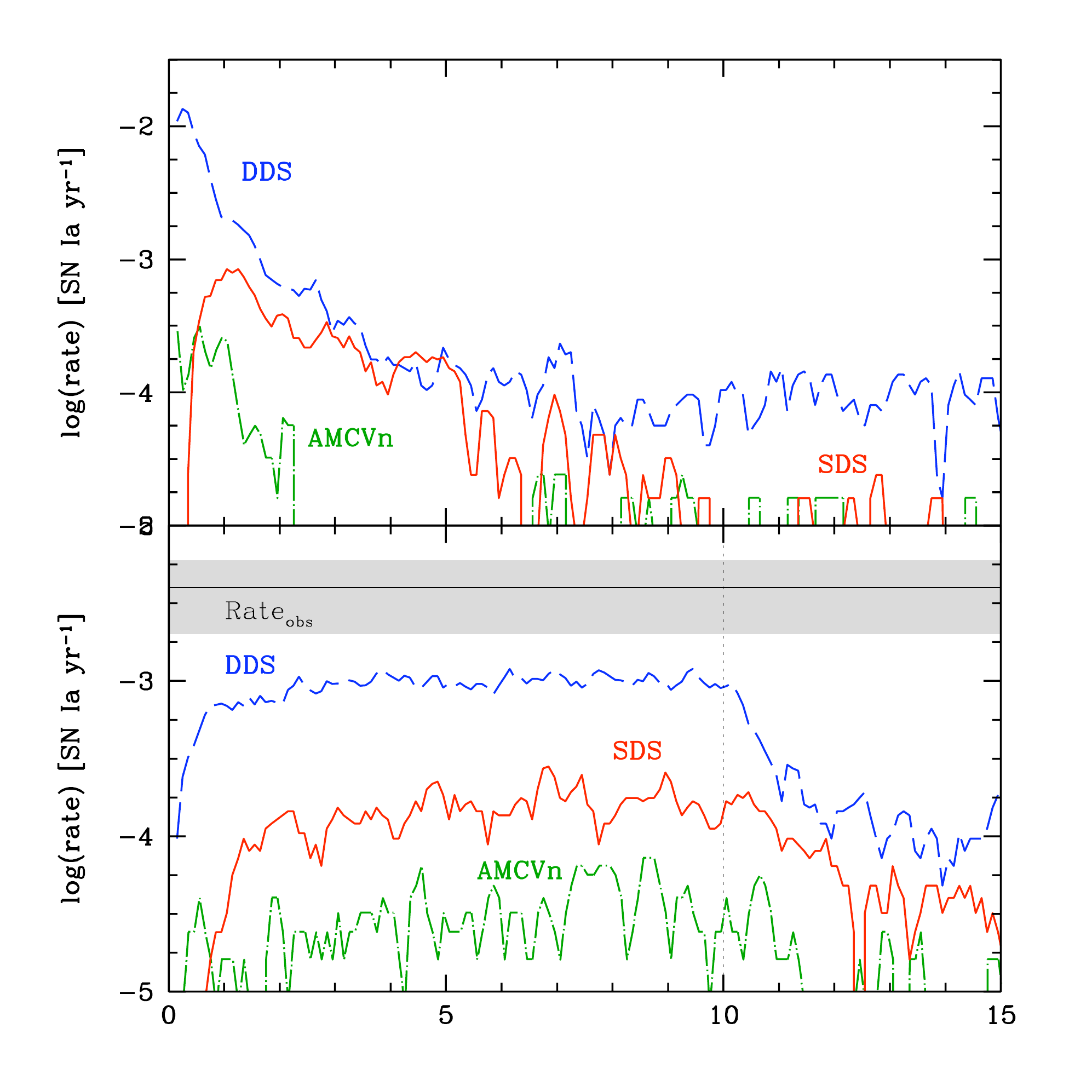}
\caption{Same as Figure 2 for the Model 2 stellar population. 
Top panel: elliptical galaxy.  
Bottom panel: spiral galaxy.  
Approximating the star formation history of the MW 
to be constant, the current Galactic SN Ia rate 
considering the DDS alone is 0.0009 yr$^{-1}$.  The combined (DDS +
SDS + AM CVn) rate is 0.001 yr$^{-1}$, which is  
below the empirical rate estimate of Cappellaro et al. (1999; $0.004 \pm 0.002$ yr$^{-1}$).}
\label{plotthree}
\end{figure*}

\end{document}